\documentclass[pdflatex]{article}
\usepackage{spconf,amsmath}
\usepackage{graphicx}
\usepackage{bm}
\usepackage{url}
\usepackage{amssymb}
\usepackage{cite}

\DeclareMathOperator{\softmax}{softmax}
\DeclareMathOperator{\concat}{concat}
\DeclareMathOperator{\lrelu}{LeakyReLU}
\DeclareMathOperator{\lin}{Linear}

\DeclareMathOperator{\clip}{clip}
\DeclareMathOperator{\CE}{CrossEntropy}
\DeclareMathOperator{\SDR}{SDR}

\title{Speech enhancement using self-adaptation and\\multi-head self-attention}

\name{Yuma Koizumi${}^{\dagger}$, Kohei Yatabe${}^{\ddagger}$, Marc Delcroix${}^{\star}$, Yoshiki Masuyama${}^{\ddagger}$, Daiki Takeuchi${}^{\ddagger}$}
\address{
${}^{\dagger}$NTT Media Intelligence Laboratories, Tokyo, Japan\\
${}^{\ddagger}$Department of Intermedia Art and Science, Waseda University, Tokyo, Japan\\
${}^{\star}$NTT Communication Science Laboratories, Kyoto, Japan
}

  \def\Hline{
  \noalign{\ifnum0=`}\fi\hrule \@height 4.\arrayrulewidth \futurelet
   \reserved@a\@xhline}

\begin{document}
\ninept
\maketitle

\begin{abstract}
This paper investigates a self-adaptation method for speech enhancement using auxiliary speaker-aware features; we extract a speaker representation used for adaptation directly from the test utterance.
Conventional studies of deep neural network (DNN)--based speech enhancement mainly focus on building a speaker independent model.
Meanwhile, in speech applications including speech recognition and synthesis, it is known that model adaptation to the target speaker improves the accuracy.
Our research question is whether a DNN for speech enhancement can be adopted to unknown speakers without any auxiliary guidance signal in test-phase.
To achieve this, we adopt multi-task learning of speech enhancement and speaker identification, and use the output of the final hidden layer of speaker identification branch as an auxiliary feature.
In addition, we use multi-head self-attention for capturing long-term dependencies in the speech and noise.
Experimental results on a public dataset show that our strategy achieves the state-of-the-art performance and also outperform conventional methods in terms of subjective quality.
\end{abstract}


\vspace{-0pt}

\section{Introduction}
\label{sec:intro}

Speech enhancement (or speech-nonspeech separation \cite{Wang_2018}) is used to recover target speech from a noisy observed signal. 
It is a fundamental task with a wide range of applications such as automatic speech recognition (ASR) \cite{NTTchime,Erdogan_2015}.
A recent advancement in this area is the use of a deep neural network (DNN) for estimating unknown parameters such as a time-frequency (T-F) mask \cite{Wang_2018}. 
In this study, we focus on DNN-based single channel speech enhancement using T-F masking; {\it i.e.} a T-F mask is estimated using a DNN and applied to the T-F represention of the observation, then the estimated signal is re-synthesized using the inverse transform.

{\it Generalization} is an important requirement in DNN-based speech enhancement to enable enhancing unknown speakers' speech.
To achieve this, several previous studies train a speaker independent DNN using many speech samples spoken by many speakers
\cite{Erdogan_2015,segan,Will_cIRM_2016,Koizumi_ICASSP_2017,Erdogan_2018_INTERSPEECH,mmsegan,dfl,Koizumi_TASL_2018,Takeuchi_2019,metricgan,Kawanaka_2020,Takeuchi_2020}.
Meanwhile, in other speech applications, model {\it specialization} to the target speaker has succeeded \cite{synth01,asr01}.
In text-to-speech synthesis (TTS), the target speaker model is trained using samples spoken by a target speaker, and that has achieved high performance \cite{synth01}.
In addition, by adapting a global ASR/TTS model to the target speaker using an auxiliary feature such as the i-vector \cite{asr01,ivec_TTS,speaker_code_TTS} and/or a speaker code \cite{speaker_code_TTS,Hojo}, ASR/TTS performance has been increased.

Success of model specialization suggests us that speaker information is important to improve the performance of speech applications including speech enhancement.
In fact, for speech separation (or multi-talker separation \cite{Wang_2018}), several works have succeeded to extract the desired speaker's speech utilizing speaker information as an auxiliary input \cite{speaker_beam,speaker_beam2,msr_speaker_profile}, 
in contrast to separating arbitrary speakers' mixture such as deep-clustering \cite{Hershey_2016} and permutation invariant training \cite{Kolbak_2017}.
A limitation of these studies is that they require a guidance signal such as adaptation utterance, because there is no way of knowing which signal in the speech-mixture is the target.
However, in speech enhancement scenario, the dominant signal is the target speech and noise is not interference speech \cite{Wang_2018}.
Thus, we consider that we can specialize a DNN for enhancing the target speech without any guidance signal in the test-phase.

In this paper, we investigate whether we can adapt a DNN to enhance the target speech while extracting speaker-related information from the observation simultaneously.
DNN-based T-F mask estimator has a speaker identification branch, which is simultaneously trained using a multi-task-learning-based loss function of speech enhancement and speaker identification.
Then, we use the output of the final hidden layer of speaker identification branch as an auxiliary feature.
In addition, to capture long-term dependencies in the speech and noise, we combine bidirectional long short-term memory (BLSTM)--based and multi-head self-attention \cite{transformer} (MHSA)--based time-series modeling.
Experimental results show that 
(i) our strategy is effective even when the target speaker is not included in the training dataset, 
(ii) the proposed method achieved the state-of-the-art performance on a public dataset \cite{dataset}, and
(iii) subjective quality was also better than conventional methods.

\section{Related works}

\begin{figure*}[ttt]
  \centering
  \includegraphics[width=175mm,clip]{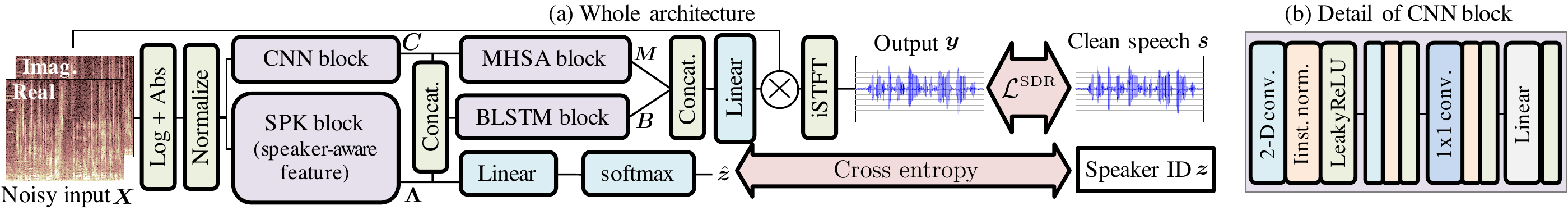}  
  \vspace{-5pt}
  \caption{Overview of proposed network architecture; (a) whole processing procedure and (b) detail of CNN block.}
  \label{fig:network}
  \vspace{-2pt}
\end{figure*}

\subsection{DNN-based speech enhancement and separation}
\label{sec:conv}

Let $T$-point-long time-domain observation $\bm{x} \in \mathbb{R}^{T}$ be a mixture of a target speech $\bm{s}$ and noise $\bm{n}$ as $\bm{x} = \bm{s} + \bm{n}$.
The goal of speech enhancement and separation is to recover $\bm{s}$ from $\bm{x}$.
In speech enhancement, $\bm{n}$ is assumed to be environmental noise and does not include interference speech signals.
Meanwhile, in speech separation, $\bm{x}$ consists of $J$ interference speech signals.

Over the last decade, the use of DNN for speech enhancement and separation has substantially advanced the state-of-the art performance by leveraging large training data.
A popular strategy is to use a DNN for estimating a T-F mask in the short-time Fourier transform (STFT)--domain\cite{Wang_2018} .
Let $\mathcal{F} : \mathbb{R}^{T} \to \mathbb{C}^{F \times K}$ be the STFT where $F$ and $K$ are the number of frequency and time bins.
The general form of DNN-based speech enhancement using T-F mask can be written as
\begin{align}
\bm{y} = 
\mathcal{F}^{\dag}\left(
\mathcal{M}(\bm{x}; \theta) \odot \mathcal{F} \left( \bm{x} \right)
\right),
\end{align}
where 
$\bm{y}$ is the estimate of $\bm{s}$, $\mathcal{F}^{\dag}$ is the inverse-STFT, $\odot$ is the element-wise product, $\mathcal{M}$ is a DNN for estimating a T-F mask, and $\theta$ is the set of its parameters.

\subsection{Auxiliary speaker-aware feature for speech separation}

An important requirement in DNN-based speech enhancement and separation is {\it generalization} that means working for any speaker.
To achieve this, in speech enhancement, several studies train a global $\mathcal{M}$ using many speech samples spoken by many speakers
\cite{Erdogan_2015,segan,Will_cIRM_2016,Koizumi_ICASSP_2017,Erdogan_2018_INTERSPEECH,mmsegan,dfl,Koizumi_TASL_2018,Takeuchi_2019,metricgan,Kawanaka_2020,Takeuchi_2020}.
Unfortunately, in speech separation, generalization cannot be achieved solely using a large scale training dataset because there is no way of knowing which signal in the speech-mixture is the target.
The most popular strategy is to separate $\bm{x}$ into $J$ speech signals, and selecting the target speech from it \cite{Hershey_2016,Kolbak_2017,deep_atractor}.

Recently, to avoid such multistage processing, the use of an auxiliary speaker-aware feature has been investigated \cite{speaker_beam,speaker_beam2,msr_speaker_profile}.
A clean speech spoken by the target speaker is also passed to the DNN.
Then, by using the clean speech as guidance, a DNN is specialized to recover the target speech.
In the SpeakerBeam method \cite{speaker_beam,speaker_beam2}, the guidance signal in the T-F-domain $\bm{A} \in \mathbb{C}^{F \times K_a}$ is converted to the sequence-summarized feature $\bm{\lambda} \in \mathbb{R}^{P}$ 
using an auxiliary neural network $\mathcal{G}: \mathbb{C}^{F \times K_a} \to \mathbb{R}^{P \times K_a}$ as
\begin{align}
\bm{\lambda} = \frac{1}{K_a} \sum_{k=1}^{K_a}  \bm{\lambda}_k, \qquad \bm{\Lambda} = ( \bm{\lambda}_1, ..., \bm{\lambda}_{K_a}) = \mathcal{G}\left( \bm{A} ; \theta_{g} \right) ,
\label{eq:seq_fet}
\end{align}
where $\theta_{g}$ is the set of parameters of $\mathcal{G}$.
Since the input of $\mathcal{G}$ is a clean speech of the target speaker, we can expect $\bm{\lambda}$ includes the speaker voice characteristics.
Thus, $\bm{\lambda}$ is used as a model-adaptation parameter by multiplying to the outputs of a hidden layer of $\mathcal{M}$.

Using auxiliary information about the target speaker has only been investigated for target speech extraction.
In this paper, we investigate it for noise reduction. 
In this case, since the noisy signal contains only speech of the target and noise, 
we expect that it would be possible to extract speaker information directly from the noisy signal and 
realize thus self-adaptation ({\it i.e.} without auxiliary guidance signal).

\section{Proposed method}
\label{sec:prop}

\subsection{Basic idea}
\label{sec:idea}

Figure\,\ref{fig:network}-(a) shows the overview of the proposed neural network.
We adopt the multi-task-learning strategy for incorporating speaker-aware feature extraction for speech enhancement.
The speech enhancement DNN has a branch for speaker identification (SPK block), and its final hidden layer's output is used as an auxiliary feature.
Both T-F mask estimation and speaker identification are trained simultaneously using a joint cost function of speech enhancement and speaker identification.

In addition, to capture the characteristics of speech and noise, not only adjacent time-frames but also long-term dependencies in a sequence should be important.
To capture longer-term dependencies, a recent research revealed that the MHSA \cite{transformer} is effective for time-series modeling in speech recognition/synthesis \cite{karita}.
Therefore, in this study, we combine BLSTM-based and MHSA-based time-series modeling.

\subsection{Implementation}
\label{sec:network}

The base architecture is a combination of a convolutional neural network (CNN) block and a BLSTM block. 
This set up is a standard architecture in DNN-based speech enhancement \cite{Erdogan_2018_INTERSPEECH}. 
We add a speaker identification branch for extracting speaker-aware auxiliary feature ({\it i.e.} SPK block) and a MHSA block to the base network.
The input of the DNN is the log-amplitude spectrogram of $\bm{X} = \mathcal{F} \left( \bm{x} \right)$, and the output of that is a complex-valued T-F mask.
Note that the input is normalized to have zero mean and unit variance for each frequency bin.
Then, the complex-valued T-F mask is multiplied to $\bm{X}$ and re-synthesized using the inverse STFT.
Hereafter, we describe the detail of each block.

{\bf CNN block:} The CNN block consists of two 2-D convolution, one 1x1 convolution, and one linear layer as shown in Fig.\,\ref{fig:network}-(b).
For both 2-D convolution layers, we used 5x5 kernel, (2,2) padding, and (1,1) stride to obtain the same size of input/output.
The number of output channel is 45 and 90 for the first and second 2-D convolution layer, respectively.
We used the instance normalization \cite{ref:IN} and leaky-ReLU activation after each CNN.
Then, CNN output is passed to the linear layer, and output $\bm{C} \in \mathbb{R}^{D \times K}$. 

{\bf SPK block:} The input feature is also passed to the SPK block.
This block consists of one CNN block and one BLSTM layer.
This CNN block consists of the same architecture of the above CNN block but the output channels of each 2-D convolution layer are 30 and 60, respectively.
Then, the CNN block output $\mathbb{R}^{D \times K}$ is passed to the BLSTM layer, then its forward and backward outputs are concatenated as $\bm{\Lambda} \in \mathbb{R}^{D \times K}$. 

{\bf BLSTM block:} Then, $\bm{C}$ and $\bm{\Lambda}$ are concatenated and passed to the BLSTM block.
The BLSTM block consists of two BLSTM layers, and its forward and backward outputs are concatenated as the output of this block $\bm{B} \in \mathbb{R}^{2D \times K}$. 
Note that, although SpeakerBeam uses the sequence-summarized feature as (\ref{eq:seq_fet}), we directly use $\bm{\Lambda}$ as a speaker-aware auxiliary feature.
Since the speaker information is captured from the noisy signal, it is possible to obtain time-dependent speaker information that could better represent the phoneme of dynamic information.

{\bf MHSA block: } As a parallel path of the BLSTM block, we use MHSA block.
This block consists of one linear layer and two cascaded MHSA modules.
First, $\bm{C}$ is passed to the linear layer to reduce its dimmension $D$ to $D/2$.
Then the linear layer output $\bm{\Gamma} \in \mathbb{R}^{D/2 \times K}$ is passed to the MHSA modules.
The input/output dimension of each MHSA module is $D/2 \times K$ thus the final output is $\bm{M} \in \mathbb{R}^{D/2 \times K}$.
For simplifying the description of this section, the detail of one MHSA module is described in Appendix A.
Note that $\bm{\Lambda}$ is not passed to this block.
The reason is we expect that this block mainly extracts long-term dependencies of noise information because speaker information is analyzed in the SPK block.
To capture non-stationary noise such as intermittent noise, the long-time similarity of the noise might be important and a speaker-aware feature and position information should not be important.

{\bf DNN outputs and loss function: } The main output of the DNN is a complex-valued T-F mask.
This mask is calculated by the last linear layer whose output dimension is $2F \times K$.
Then, the output is split into two $F \times K$ matrices, and used as the real- and imaginary-part of a complex-valued T-F mask.
In the training phase, $\bm{\Lambda}$ is also passed to a linear layer and we obtain $\bm{Z} = ( \bm{z}_1, ..., \bm{z}_{K}) \in \mathbb{R}^{L \times K}$ where $L$ is the number of speakers included in training dataset.
Then, speaker ID of $\bm{X}$ is estimated as 
$\hat{\bm{z}} = \softmax ( K^{-1} \sum_{k=1}^{K}  \bm{z}_k)$.
We use a multi-task loss, which consists of a SDR-based loss and the cross-entropy loss are calculated as
\begin{align}
\mathcal{L} &= \mathcal{L}^{\mbox{\tiny SDR}} + \alpha \CE ( \bm{z}, \hat{\bm{z}} ),\\
\mathcal{L}^{\mbox{\tiny SDR}} &= 
- \frac{1}{2}
\left(
\clip_{\beta}\left[
\SDR ( \bm{s}, \bm{y} )
\right]
+
\clip_{\beta}\left[
\SDR ( \bm{n}, \bm{m} )
\right]
\right),
\end{align}
where 
$\SDR ( \bm{s},\bm{y} ) = 
10 \log_{10} \left( \lVert \bm{s} \rVert _2^2  / \lVert \bm{s} - \bm{y} \rVert _2^2\right)
$,
$\lVert \cdot \rVert_2$ is $\ell_2$ norm,
$\bm{m} = \bm{x} - \bm{y}$, 
$\alpha > 0$ is a mixing parameter, 
$\clip_{\beta}[x] = \beta \cdot \tanh(x / \beta)$, 
$\beta > 0$ is a clipping parameter \cite{Erdogan_2018_INTERSPEECH}, and $\bm{z}$ is the true speaker label $\bm{X}$.

\section{Experiments}
\label{sec:exp}

To investigate whether our strategy is effective for DNN-based speech enhancement, we conducted three experiments; (i) a verification experiment using a small dataset, (ii) an objective experiment using a public dataset, and (iii) a subjective experiment.
The purposes of each experiment were (i) verifying the effectiveness of auxiliary feature, (ii) comparing the performance with conventional studies, and (iii) evaluating not only objective metrics but also subjective quality, respectively.
In all experiments, we utilized the VoiceBank-DEMAND dataset constructed by Valentini {\it et al.} \cite{dataset} 
which is openly available and frequently used in the literature of DNN-based speech enhancement \cite{segan,mmsegan,dfl,metricgan}.
The train and test sets consists of 28 and 2 speakers (11572 and 824 utterances), respectively.

\subsection{Verification experiments}
\label{sec:exp_varif}

First, we conducted a verification experiment to investigate the effectiveness of auxiliary speaker-aware feature.
Since this experiment mainly focuses on the effectiveness of the SPK block, we modified the DNN architecture illustrated in Fig.\,\ref{fig:network}.
We removed all CNNs and MHSA modules, and all BLSTMs were chenged to the bidirectional-gated recurrent unit (BiGRU) with one hidden layer.
The unit size of the BiGRU was $D=200$.
In this experiment, one of the speakers in the training dataset \url{p226} was used as the target speaker.
In order to use the same number of training samples for each speaker, the training dataset was separated into $300\times28$ training samples and other samples spoken by each speaker were used for training.
Then, we tested three types of training as follows:
\vspace{-2pt}
\begin{description}
  \setlength{\parskip}{0cm} 
  \setlength{\itemsep}{0cm} 
\item[{\tt Close: }] Trained using the target speaker's samples, that is, the speaker dependent model. This score indicates the ideal performance and just for reference because it cannot be used in practice.
\item[{\tt Open: }] Trained using other 27 speakers' samples, that is, the speaker independent model (the scope of conventional studies).
\item[{\tt Open+SPK: }] Trained using other 27 speaker's samples, and the DNN has a SPK branch. The output of SPK block was used as an auxiliary speaker-aware feature, that is, self-adaptation model (the scope of this study).
\end{description}
\vspace{-2pt}
Thus, for {\tt Close}, 300 samples spoken by \url{p226} were the training data, and for {\tt Open} and {\tt Open+SPK}, $300\times 27$ samples except \url{p226} were training data.
The number of test samples was 53.
In addition, we used data augmentation by swapping noise of randomly selected two samples.
The training setup and STFT parameters were the same as the objective experiment described in Sec.\,\ref{sec:obj_eval}.

\begin{table}[ttt]
\centering
\caption{Results of verification experiment.} 
\begin{tabular}{l|ccccc} \hline
	Method 				&	SI-SDR 		& PESQ			&	CSIG    		&	CBAK			& COVL	\\ \hline
	{\tt Noisy}				&	$5.96$		& $1.56$			&	$2.44$		&	$2.07$			& $1.96$  \\ 
	{\tt Close}				&	${\bf 14.59}$	& ${\bf 2.18}$		&	${\bf 3.10}$	&	${\bf 2.84}$		& ${\bf 2.59}$  \\
	{\tt Open}				&	$14.28$		& $2.11$			&	$2.99$		&	$2.79$			& $2.50$  \\ 
	{\tt Open+SPK}		& 	$14.48$		& $2.15$			&	$2.96$		& 	$2.82$			& $2.51$   \\ \hline
\end{tabular}
\label{tbl:vel_score}
\end{table}

We used the perceptual evaluation of speech quality (PESQ), CSIG, CBLK, and COVL as the performance metrics which are the standard metrics for this dataset \cite{segan,mmsegan,dfl,metricgan}.
The three composite measures CSIG, CBAK, and COVL are the popular predictor of the mean opinion score (MOS) of the target signal distortion, background noise interference, and overall speech quality, respectively \cite{measure}.
In addition, as the standard metric in speech enhancement, we also evaluated scale-invariant SDR (SI-SDR) \cite{SISDR}.
Table\,\ref{tbl:vel_score} shows the experimental results.
By comparing {\tt Open} and {\tt Open+SPK}, scores of {\tt Open+SPK} were better than {\tt Open} except CSIG, and got close to {\tt Close}'s upper bound score.
This result suggest us the effectiveness of auxiliary speaker feature extracted by SPK block for DNN-based speech enhancement.

\subsection{Objective evaluations}
\label{sec:obj_eval}

We evaluated the DNN described in Sec.\,\ref{sec:network} ({\tt Ours}) on the same dataset and metrics of conventional studies \cite{segan,mmsegan,dfl,metricgan}, {\it i.e.} using all training data and evaluated on PESQ, CSIG, CBLK, and COVL.
The output size of CNN block was $D=600$ and the number of heads of each MHSA module was $H=4$.
The mixing and clipping parameters were the same as the verification experiment.
To evaluate effectiveness of SPK block and MHSA block, we also evaluated only adding either one block;
{\tt MHSA} w/o SPK block and {\tt SPK} w/o MHSA block.
The swapping-based data augmentation was used which is described in Sec.\,\ref{sec:exp_varif}.
We fixed the learning rate for the initial 100 epochs and decrease it linearly between 100--200 epochs down to a factor of 100 using ADAM optimizer, where we started with a learning rate of 0.001. 
We always concluded training after 200 epochs.
The STFT parameters were a frame shift of 128 samples, a DFT size of 512, and a window was 512 points the Blackman window.

The proposed method was compared with 
speech enhancement generative adversarial network (SEGAN) \cite{segan},
MMSE-GAN \cite{mmsegan},
deep feature loss (DFL) \cite{dfl},
and 
Metric-GAN \cite{metricgan},
because these methods have been evaluated on the same dataset.
Table\,\ref{tbl:score} shows the experimental results.
As we can see from this table, the proposed method ({\tt Ours}) achieved better performance than conventional methods in all metrics.
In addition, we observed from the attention map that the attention module tends to pay attention to longer context when there was impulsive noise or consonants.
This may help reduce noise and be the reason of that  {\tt MHSA} achieved the best CBAK score.

\begin{table}[ttt]
  \vspace{-10pt}
\centering
\caption{Objective evaluation results on public dataset \cite{dataset}.} 
\begin{tabular}{l|ccccc} \hline
	Method 						&	 PESQ			&	CSIG   	 		&	CBAK			& COVL		\\ \hline
	{\tt Noisy}						&	$1.97$			&	$ 3.35$			&	$ 2.44$			& $2.63$ \\
	{\tt SEGAN} \cite{segan}		&	$2.16$			&	$ 3.48$			&	$ 2.94$			& $2.80$ \\ 
	{\tt MMSE-GAN} \cite{mmsegan}	& 	$2.53$			&	$ 3.80$			& 	$3.12$			& $3.14$ \\
	{\tt DFL} \cite{dfl}				&	n/a				&	$ 3.86$			&	$ 3.33$			& $3.22$ \\
	{\tt MetricGAN} \cite{metricgan} 	&	$2.86$			&	$ 3.99$			&	$ 3.18$			& $3.42$ \\ \hline
	{\tt MHSA}						&	$2.93$			&	$4.10$			&	${\bf 3.46}$		& $3.51$\\ 
	{\tt SPK}						&	$2.95$			&	$4.10$			&	$3.41$			& $3.52$\\ 
	{\tt Ours}						&	${\bf 2.99}$		&	${\bf 4.15}$		&	$3.42$ 			& ${\bf 3.57}$ \\ \hline
\end{tabular}
\label{tbl:score}
\end{table}

\begin{figure}[ttt]
  \centering 
  \includegraphics[width=85mm,clip]{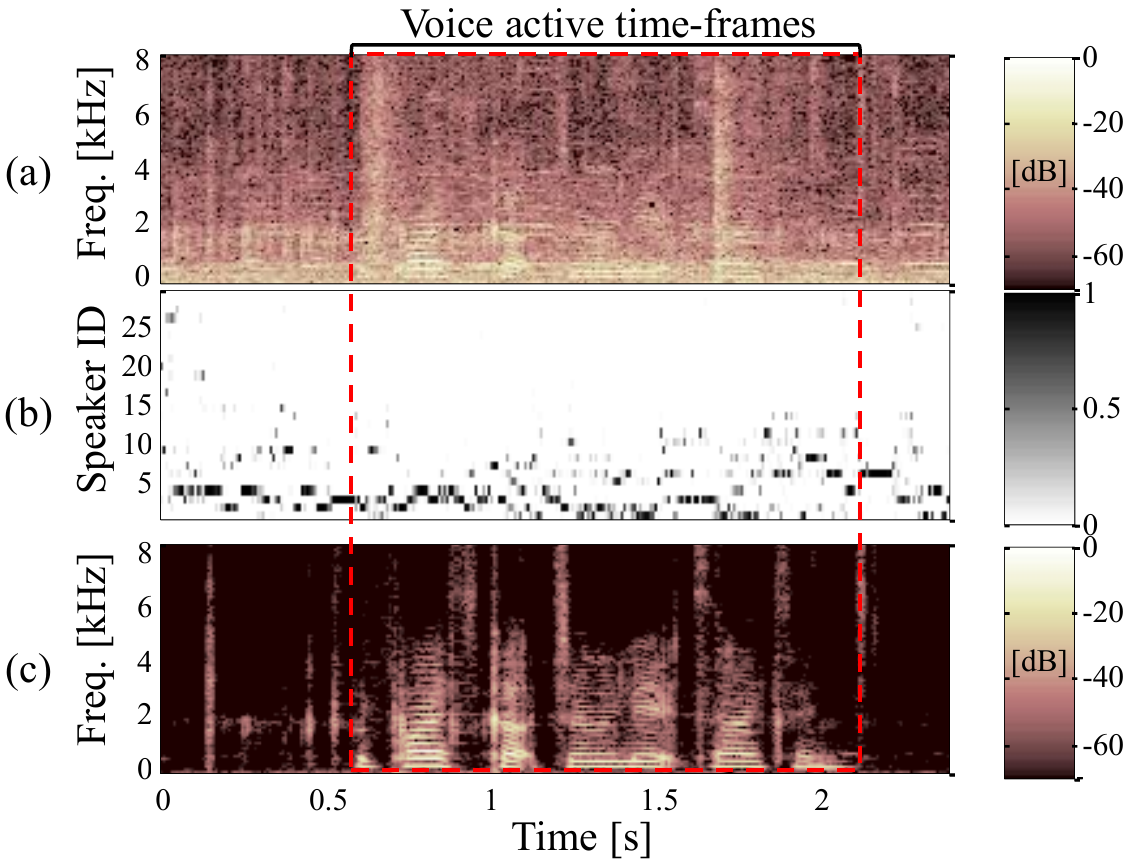}  
  \vspace{-5pt}
  \caption{Example of speech enhancement result.
  Each figure shows
  (a) spectrogram of input,
  (b) speaker identification result on each time-frame where speaker IDs have been sorted by descending order of frequency of appearance, and 
  (c) spectrogram of output, respectively.
  }
  \label{fig:result_exp}
  \vspace{-2pt}
\end{figure}

Figure\,\ref{fig:result_exp} shows the speech enhancement result of \url{p257_070.wav} which is a test sample spoken by a female speaker under a low signal-to-noise ratio (SNR) condition.
The top and bottom figures show the spectrograms of the input and output, respectively.
Figure\,\ref{fig:result_exp}-(b) shows speaker identification result on each time-frame which is calculated for visualization as $\hat{\bm{z}}_k = \softmax ( \bm{z}_k )$ instead of $\hat{\bm{z}}$.
Although the SPK branch estimated the speaker of the whole utterance was \url{p228} (a female) as $\hat{\bm{z}}$, the middle figure ({\it i.e.}\,$\hat{\bm{z}}_k$) shows the SPK branch might select different speaker frame-by-frame.
Actually, in the voice active time-frames of this utterance, 64\% and 84\% of the time-frames were occupied by 3 and 5 speakers, respectively; \url{p268} (id: 1), \url{p228} (id: 2), \url{p256} (id: 3), \url{p239} (id: 4), and \url{p258} (id: 5).
\url{p256} is a male speaker whose voice is a bit hoarse, and the male speaker is mainly selected at consonant time-frames.
Since the SPK block can extract speaker-aware information,
{\tt SPK} achieved better PESQ and COVL score than {\tt MHSA}, and its convention {\tt Ours} achieved the state-of-the-art performance on a public dataset for speech enhancement.

\subsection{Subjective evaluations}
\label{sec:sub_eval}

\begin{figure}[ttt]
  \centering
  \includegraphics[width=70mm,clip]{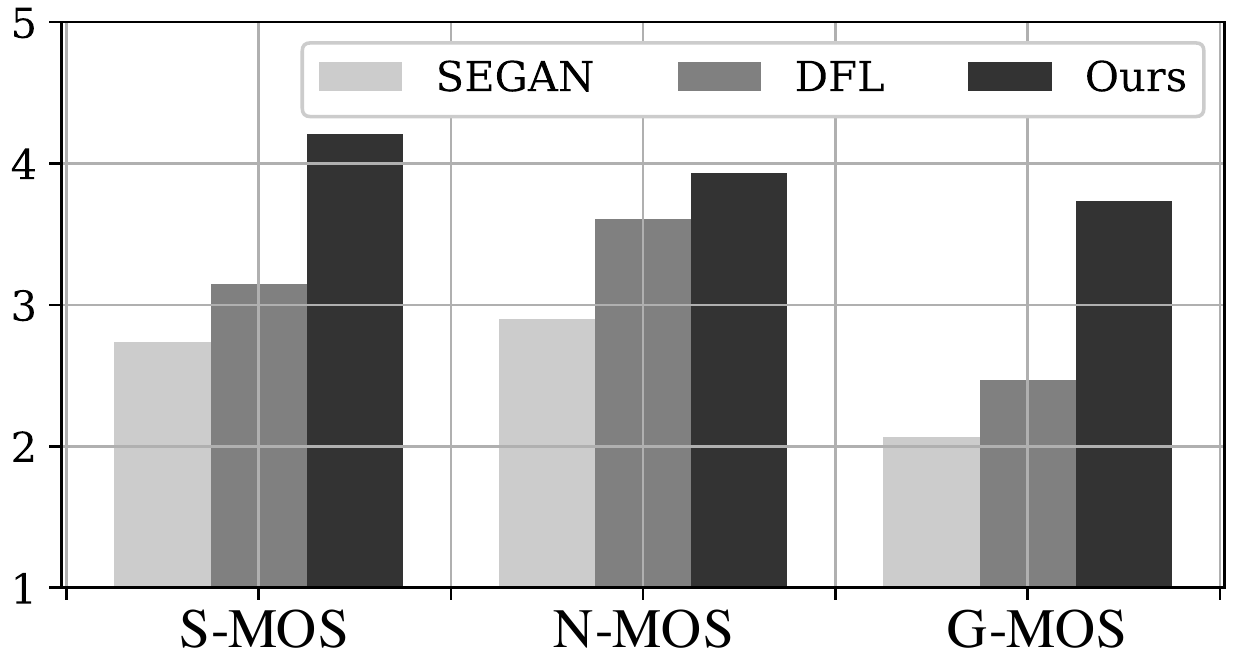}  
  \vspace{-5pt}
  \caption{Subjective evaluation results according to ITU-T P.835.}
  \label{fig:mos}
  \vspace{-2pt}
\end{figure}

We conducted a subjective experiment.
The proposed method was compared with
SEGAN \cite{segan}
and 
DFL \cite{dfl} 
because speech samples of both methods are openly available in DFL's web-page \cite{dfl_web}.
We selected 15 samples from Tranche 1--3 data from the web-page (low SNR conditions).
The speech samples of the proposed method used in this test are also openly available\footnote{\url{https://sites.google.com/site/yumakoizumiweb/publication/icassp2020}}.
The subjective test was designed according to ITU-T P.835 \cite{P835}.
In the tests, the participants rated three different factors in the samples:
speech mean-opinion-score (S-MOS),
subjective noise MOS (N-MOS),
and overall MOS (G-MOS).
Ten participants evaluated the sound quality of the output signals.
Figure\,\ref{fig:mos} shows the results of the subjective test.
For all factors, the proposed method achieved the highest score, and statistically significant differences were observed in a paired one sided $t$-test ($p < 0.01$).
From these results, it is suggested that the SPK block can extract speaker-aware information and it is effective for DNN-based speech enhancement.

\section{Conclusions}
\label{sec:cncl}

We investigated the use of self-adaptation for DNN-based speech enhancement; we extracted a speaker representation used for adaptation directly from the test utterance.
A multi-task-based cost function was used for simultaneously training DNN-based T-F mask estimator and speaker identifier for extracting a speaker representation.
Three experiments showed that 
(i) our strategy was effective even if the target speaker is unknown, 
(ii) the proposed method achieved the state-of-the-art performance on a public dataset \cite{dataset}, and
(iii) subjective quality was also better than conventional methods.
Thus, we concluded that self-adaptation using speaker-aware feature is effective for DNN-based speech enhancement.

\begin{center}
{\bf Appendix A. \ \ Detail of MHSA module}
\end{center}

Here, we briefly describe the detail of MHSA module \cite{transformer}.
Let $H$ be the number of heads in MHSA.
First, $\bm{\Gamma}$ is inputted to the layer normalization.
Then, $h$-th head's attention matrix $\bm{A}_h$ is calculated as 
$\bm{A}_h = \mbox{softmax} ( d^{-1/2} \cdot \tilde{\bm{A}}_h )$
where $d = D/(2H)$ is a scaling parameter, and
\begin{align}
\tilde{\bm{A}}_h = \left( \bm{W}_{h,q} \bm{\Gamma} \right)^{\mathsf{T}} \bm{W}_{h,k} \bm{\Gamma} \in \mathbb{R}^{K \times K}.
\end{align}
Here ${}^{\mathsf{T}}$ is the transpose.
The size of $\bm{W}_{h,q}$ and $\bm{W}_{h,k}$ are $ D/(2H) \times D/2 $.
Then, $h$-th head's context matrix is calculated as
\begin{align}
\bm{E}_h = \bm{A}_h \left( \bm{W}_{h,v} \bm{\Gamma} \right)^{\mathsf{T}} \in \mathbb{R}^{T \times D/(2H)},
\end{align}
where the size of $\bm{W}_{h,v}$ is also $ D/(2H) \times D/2 $.
Then, $\bm{W}_p \in \mathbb{R}^{D/2 \times D/2}$ is multiplied to the concatenated $\bm{E}_h$ and added to $\bm{\Gamma}$ as
\begin{align}
\bm{E} = \left( \concat [ \{ \bm{E}_h \}_{h=1}^{H} ] \bm{W}_p \right)^{\mathsf{T}}
+ \bm{\Gamma} \in \mathbb{R}^{D/2 \times K}.
\end{align}
Finally, $\bm{E}$ is passed to the second layer normalization,
and passed to the point-wise feed-forward layer as
\begin{align}
\bm{M} = 
\lin^{(1)} \left( 
\lrelu \left( 
\lin^{(2)} \left( 
\bm{E}
\right)
\right)
\right),
\end{align}
where $\lin(\cdot)$ denotes feed-forward NN, and the output size of $\lin^{(1)}$ and $\lin^{(2)}$ are $D/2$ and $3D/2$, respectively.

\clearpage
\bibliographystyle{IEEEbib}
\bibliography{refs}

\end{document}